\documentclass[%
 reprint,
 superscriptaddress,
 amsmath,amssymb,
prl
]{revtex4-1}

\usepackage{graphicx}
\usepackage{color}
\usepackage{dcolumn}
\usepackage{bm}

\begin{document}

\preprint{APS/123-QED}

\title{Muonium state exchange dynamics in n-type Gallium Arsenide}

\author{K.~Yokoyama}
\email{koji.yokoyama@stfc.ac.uk}
\affiliation{
ISIS Neutron and Muon Facility, STFC Rutherford Appleton Laboratory, Didcot, OX11 0QX, United Kingdom
}
\author{J.~S.~Lord}
\affiliation{
ISIS Neutron and Muon Facility, STFC Rutherford Appleton Laboratory, Didcot, OX11 0QX, United Kingdom
}
\author{P.~W.~Mengyan}
\affiliation{
Department of Physics, Northern Michigan University, Marquette, Michigan 49855, USA
}
\author{M.~R.~Goeks}
\affiliation{
Department of Physics, Northern Michigan University, Marquette, Michigan 49855, USA
}
\author{R.~L.~Lichti}
\affiliation{
Department of Physics and Astronomy, Texas Tech University, Lubbock, Texas 79409-1051, USA
}

\date{\today}

\begin{abstract}
Muonium (Mu), a pseudo-isotope atom of hydrogen with a positively charged muon at the place of the proton, can form in a wide range of semiconductor materials.
They can appear in different states, depending on its charge state and microscopic site within a crystal lattice.
After the Mu formation, they undergo interactions with free charge carriers, electronic spins, and other Mu sites, and form a dynamic network of state exchange.
We identified the model of Mu dynamics in n-type Gallium Arsenide using the density matrix simulation and photoexcited muon spin spectroscopy technique.
Fitting to the dark and illuminated $\mu$SR data provided transition rates between Mu states, which in turn showed the underlying mechanism of the $\mu$SR time spectra.
Deduced capture/scattering cross sections of the Mu states reflected microscopic dynamics of Mu.
Illumination studies enable us to measure interactions between Mu and generated minority carriers, which are unavailable in dark measurements.
The methodology we developed in this study can be applied to other semiconductor systems for a deeper microscopic understanding of the Mu state exchange dynamics.
\end{abstract}

\maketitle

Hydrogen (H) is an important impurity in semiconductor engineering and device processing because of its fast lattice diffusion, high reactivity, and ubiquity in fabrication processes.
In semiconductor materials, hydrogen can take several different charge states and modify host's electrical and optical properties.
They are beneficial in some cases ({\it e.g.} passivating dangling bonds in bulk and on surfaces), and harmful in other cases ({\it e.g.} passivating intentional dopant atoms) \cite{Myers}.
Therefore, it is important to acquire comprehensive knowledge of H behaviors in electronic materials.
However, because of the elusive nature of hydrogen, only a few experimental techniques are available for the study, such as vibrational spectroscopy \cite{Stavola} and deep level transient spectroscopy \cite{Peaker}.
In particular, investigating isolated H atoms can be a difficult task because they are often in a precursor state that rapidly react and form a stable final state.
Muon spin spectroscopy, collectively known as $\mu$SR (muon spin relaxation, rotation, and resonance), has made significant contribution in this avenue by clarifying the electronic structure and dynamics of isolated muonium (Mu) states \cite{Patterson, Cox_Rev, Chow_Rev}.
Since a positively charged muon $\mu^+$ has a 1/9-th of the proton mass and a mean lifetime of 2.2~$\mu$s,
its bound state, Mu$^0$ = $\mu^+$~+~e$^-$, is considered as a light, unstable pseudo-isotope atom of hydrogen, which can mimic the H states in materials.

There are two types of paramagnetic Mu$^0$ states commonly found in semiconductors, which are characterized by their hyperfine (HF) coupling constant \cite{Patterson, Cox_Rev, Chow_Rev}:
(1) Mu$_{\text{T}}^0$, a tetrahedral interstitial Mu with a large isotropic HF constant A$_\mu$ (in the order of GHz), and
(2) Mu$_{\text{BC}}^0$, a bond center Mu with an anisotropic HF constant axially symmetric about a $<$111$>$ crystal axis.
For Mu$_{\text{BC}}^0$, its HF constants are characterized by A$_\parallel$ and A$_\perp$, representing components parallel and perpendicular to the $<$111$>$ axis respectively, with their magnitudes roughly an order of magnitude smaller than A$_\mu$.
Their microscopic behaviors are qualitatively similar from one semiconductor to another, where Mu$_{\text{T}}^0$ is in a delocalized Block state and diffuses rapidly, and Mu$_{\text{BC}}^0$ is stationary on the timescale of the muon lifetime.
Just like hydrogen in many semiconductors \cite{VandeWalle}, muonium also shows compensating properties and can exist as Mu$^+$ (a bare $\mu^+$) in p-type and Mu$^-$ ($\mu^+$~+~2~e$^-$) in n-type materials under equilibrium conditions \cite{Lichti_PRL, Lichti_PRB}.
It is widely known that there are four types of isolated Mu centers in semiconductors:
Mu$_{\text{T}}^0$, Mu$_{\text{BC}}^0$, Mu$_{\text{T}}^-$, and Mu$_{\text{BC}}^+$.
Furthermore, transitions can take place between the centers via charge carrier exchange, spin exchange, and site change interaction, which can result in a formation of a dynamic cyclic network \cite{Chow_Rev, Lichti_Rev}.
This mechanism was extensively studied in Si using the radio frequency (RF) $\mu$SR technique, where diamagnetic states are selectively measured in resonant conditions \cite{Kreitzman, Scheuermann_RF}.
More recently, the photoexcited $\mu$SR (photo-$\mu$SR) technique was utilized to optically pump Si wafers and study the transition dynamics enhanced by generated excess charge carriers \cite{Kadono_Si_PRL, Scheuermann_photo, Fan, Yokoyama_PRL}.
The optical method at pulsed sources enables us to generate excess carriers without heating samples, and thus keep band gap energies the same.
Some of the recent studies applied density matrix methods to calculate time evolution of a muon spin ensemble in the cyclic models, and gave a rate for each transition channel \cite{Fan, Yokoyama_PRL}.
If a state exchange involves a free carrier ({\it e.g.} Mu$_{\text{T}}^0$ + e$^-$ $\rightarrow$ Mu$_{\text{T}}^-$), its capture cross section can be calculated from an obtained transition rate (see below).
Therefore, these measurements provide valuable insights into interactions between isolated H atoms and charge carriers, which are otherwise extremely difficult to study.

In this Letter, we report both dark and illuminated $\mu$SR experiments on n-type GaAs (n-GaAs) and elucidate the Mu transition dynamics using the density matrix method.
Generally, muon signals from semiconductors are a convolution of Mu state exchange dynamics;
the analysis presented here can deconvolute it into each component and help us understand background mechanisms.
We also demonstrate that illumination by monochromatic laser light provides precise control of an excess carrier density $\Delta$n;
this will allow us to excite some of transition channels in the cyclic models, which are inaccessible in dark measurements.
The discovered model in n-GaAs turns out to be simpler than the case of Si (or Ge), and can be considered as a basic system of isolated Mu undertaking exchange interactions with free carriers.

\begin{figure}
\includegraphics{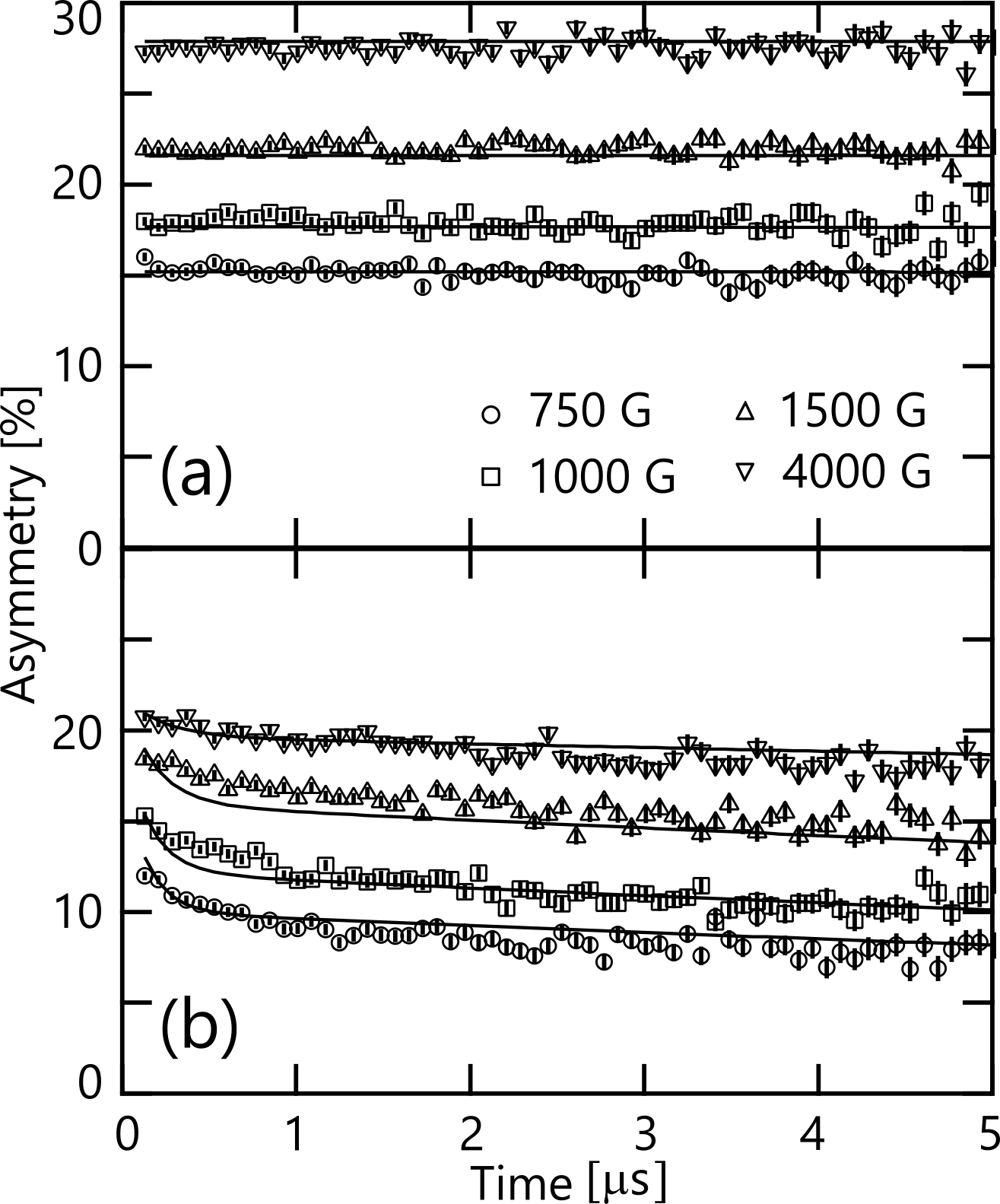}
\caption{\label{fig:Fig1_dark_spectra}
Muon asymmetry time spectra measured in dark at (a) 290~K and (b) 20~K under four representative fields.
Five million muon decay events were averaged for each spectrum.
The full muon asymmetry in HIFI is about 28~\%.
Solid lines denote the fit result (see text).
}
\end{figure}

The experiment was carried out using the high magnetic field (HIFI) muon spectrometer \cite{Lord_HIFI} at the ISIS Pulsed Neutron and Muon Source at the STFC Rutherford Appleton Laboratory in the UK \cite{DOI}.
A detailed discussion of the illumination facility and associated equipment can be found elsewhere \cite{Yokoyama_RSI}.
In this experiment, the sample was a single crystal, Si-doped n-GaAs wafer with an average carrier concentration of N$_\text{d}$ = 2.5 $\times$ 10$^{16}$~cm$^{-3}$.
The wafer had a diameter of 50~mm and a thickness of 450~$\mu$m, with the $<$100$>$ crystal axis perpendicular to the surface.
The wafer was single-side polished; laser light illuminated the polished side, and any transmitted light was scattered and diffused on the other side, where the surface was chemically etched.
The sample was contained in a helium-gas purged sample cell mounted on a closed cycle refrigerator.
The gas purged construction ensured uniform and mechanical stress-free cooling.
On the sample cell, muons are implanted from one side through a titanium window (100~$\mu$m thick), whereas laser light, propagating in the opposite direction, illuminated the sample from the other side through a glass window \cite{Yokoyama_RSI}.
Aluminum sheet degraders were used to adjust the implantation depth of ``surface'' muons (with a kinetic enegy of 4~MeV), such that the muon distribution profile was centered in the wafer \cite{Yokoyama_APL}.

Firstly, we carried out measurements in dark conditions to find a single model that can describe $\mu$SR time spectra in the temperature range.
Figure \ref{fig:Fig1_dark_spectra}(a) and (b) show the $\mu$SR spectra at 290 and 20~K respectively under four representative longitudinal fields (LF), where the field vector is parallel to the initial muon spin.
At 290~K, the time spectra show no relaxation, and a muon spin repolarization is observable as the field increases.
Kadono et al. attributed this to a precursor Mu state, presumably Mu$_{\text{T}}^0$, rapidly converted to the stable Mu$_{\text{T}}^-$ state upon capturing a free electron \cite{Kadono_GaAs_1994}.
The time spectra appear to be flat because spin relaxation in the precursor state and the subsequent state conversion are too fast to observe at a pulsed muon facility.
In contrast, a relaxing component appears at 20~K; they attributed this to spin relaxation in the Mu$_{\text{BC}}^0$ state, which was considered to be more predominant in low temperatures because of the metastability of the Mu$_{\text{BC}}^0$ center in GaAs.
As the temperature (T) rises, the transition rate from Mu$_{\text{BC}}^0$ to Mu$_{\text{T}}^0$ increases \cite{Kadono_GaAs_1994}.

\begin{figure}
\includegraphics{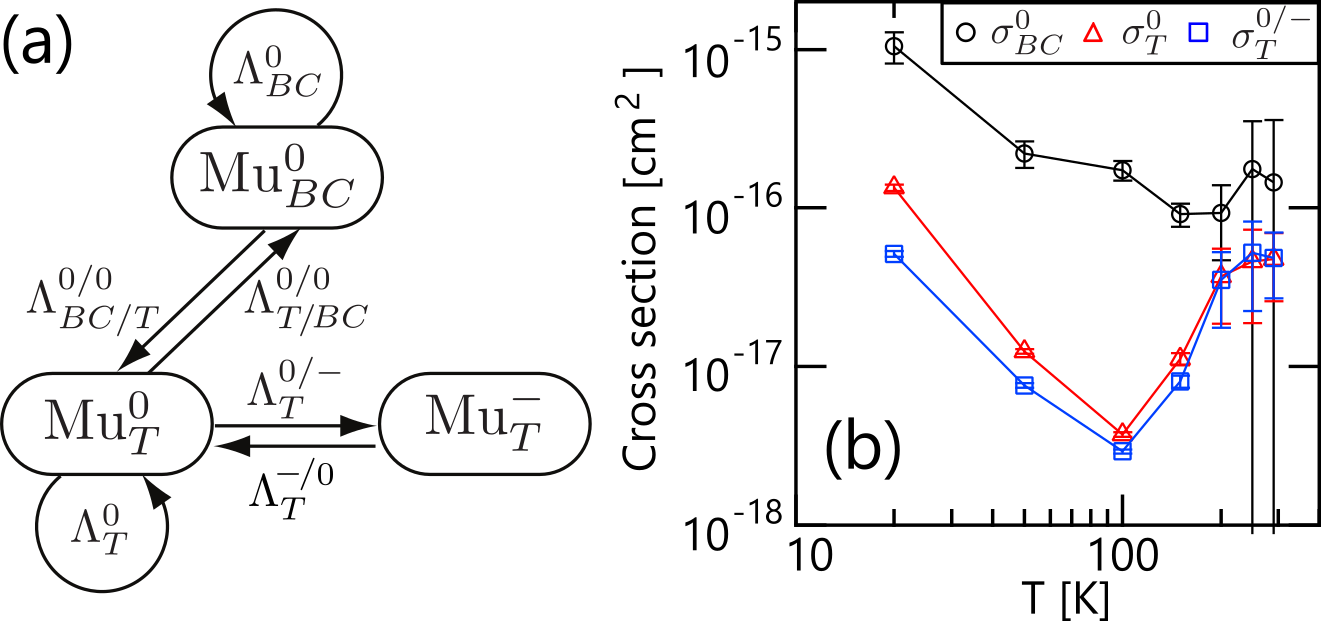}
\caption{\label{fig:Fig2_Mu_diagrams}
(a) The model of Mu state exchange in dark n-GaAs.
A superscript and subscript of $\Lambda$ indicate the charge state and Mu site respectively, with a slash between before and after a transition.
(b) Temperature dependence of scattering/capture cross sections. Straight lines connecting points are a guide to the eye.
}
\end{figure}

On the basis of the previous works, we developed a model shown in FIG.~\ref{fig:Fig2_Mu_diagrams}(a), where three Mu centers (Mu$_{\text{T}}^0$, Mu$_{\text{BC}}^0$, and Mu$_{\text{T}}^-$) exchange the states.
In this model, transitions between Mu$_{\text{T}}^0$ and Mu$_{\text{T}}^-$ require capturing a free charge carrier:
Mu$_{\text{T}}^0$ + e$^-$ $\rightarrow$ Mu$_{\text{T}}^-$ and 
Mu$_{\text{T}}^-$ + h$^+$ $\rightarrow$ Mu$_{\text{T}}^0$
(or releasing an electron spontaneously, Mu$_{\text{T}}^-$ $\rightarrow$ Mu$_{\text{T}}^0$ + e$^-$).
The site exchange interaction between Mu$_{\text{BC}}^0$ and Mu$_{\text{T}}^0$ requires phonons to overcome the energy barrier.
For the Mu$_{\text{BC}}^0$ and Mu$_{\text{T}}^0$ center, spin exchange interaction can take place between a bound and conduction electron, resulting in a muon spin relaxation due to HF interaction \cite{Yokoyama_APL}.
Positively charged Mu$_{\text{BC}}^+$ states are energetically unfavorable in n-type samples and not included in the model.
Each state exchange is characterized by a transition rate $\Lambda$ as shown in FIG.~\ref{fig:Fig2_Mu_diagrams}(a).
Similarly, the spin exchange interaction in Mu$_{\text{BC}}^0$ and Mu$_{\text{T}}^0$ are characterized by $\Lambda_{\text{BC}}^0$ and $\Lambda_{\text{T}}^0$ respectively.
Density matrix simulations on this model were applied to fit the the $\mu$SR time spectra.

The simulation was carried out using QUANTUM, a Python program to solve the time evolution of the muon spin using the density matrix method \cite{Lord_Quantum}.
Representative source codes and details of the fitting procedures can be found in Supplemental Material \cite{SM}.
Briefly, the code simulates the transition dynamics taking place after an initial distribution of the Mu states.
The HF constants, A$_\mu$ = 2.88~GHz, A$_\parallel$ = 217.8~MHz, and A$_\perp$ = 87.7~MHz, were adapted from Ref. \cite{Patterson}.
All transition rates are defined in the code, and $\Lambda_{\text{BC}}^0$ and $\Lambda_{\text{T}}^0$ are described as a relaxation rate of a bound electronic spin on each Mu.
After a calculation, the program returns a normalized muon spin asymmetry time spectrum (hence, this is a unity at t = 0), which is then multiplied by a scaling factor to match experimental data.
The scaling factor is determined from the 290~K data because the relaxing Mu$_{\text{BC}}^0$ component is negligibly small at RT and the muon signal is entirely diamagnetic.
At each T, the measurement was carried for approximately 15 fields ranging between 300 and 4000~G.
To avoid complications, we did not include low-LF data in the simulation, where level-crossing resonances induced by quadruple moments of Ga and As nuclei can cause relaxations \cite{Chow_PRB}.
To improve the fitting accuracy, a global fit was performed on this set of LF-scan data.
Note that a fractional yield of Mu$_{\text{T}}^0$ upon muon implantation, $f(\text{Mu}_{\text{T}}^0)$, is a fit parameter, but $f(\text{Mu}_{\text{BC}}^0)$ is a dependent variable and determined by $f(\text{Mu}_{\text{BC}}^0) = 1 - f(\text{Mu}_{\text{T}}^0)$.

\begin{table*}
\centering
\caption{Results of global fitting to the dark, LF-scan data measured at 290 and 20~K. The unit of $\Lambda$'s is in MHz.}
\label{table:dark_results}
\begin{tabular}{|c|c|c|c|c|c|c|c|}
\hline
T [K] & $f(\text{Mu}_{\text{T}}^0)$ & $\Lambda_{\text{T}}^0$ & $\Lambda_{\text{T}}^{0/-}$ & $\Lambda_{\text{T}}^{-/0}$ & $\Lambda_{\text{BC}}^0$ & $\Lambda_{\text{BC/T}}^{0/0}$ & $\Lambda_{\text{T/BC}}^{0/0}$\\ \hline
290 & 0.99 $\pm$ 0.06 & 49 $\pm$ 22  & 50  $\pm$ 22 & 0.000  $\pm$ 0.001 & 147  $\pm$ 216 & 38 $\pm$ 55 & 0.0  $\pm$ 0.7 \\ \hline
20 & 0.66 $\pm$ 0.02 & 10.0 $\pm$ 0.5 & 3.8  $\pm$ 0.2 & 0.074 $\pm$ 0.002 & 77  $\pm$ 17 & 24  $\pm$ 7 & 0.07 $\pm$ 0.01 \\ \hline
\end{tabular}
\end{table*}

Obtained fit results in Table~\ref{table:dark_results} show us background mechanisms of the $\mu$SR spectra in FIG.~\ref{fig:Fig1_dark_spectra}.
At 290~K, all muons are initially in the Mu$_{\text{T}}^0$ state ($f(\text{Mu}_{\text{T}}^0)$ $\simeq$ 1), which are subsequently rapidly depolarized via the spin exchange interaction (49~MHz) or converted to Mu$_{\text{T}}^-$ (50~MHz).
However, the backward transition, Mu$_{\text{T}}^-$ $\rightarrow$ Mu$_{\text{T}}^0$, rarely happens in this electron-rich environment.
This behavior explains the repolarization by external fields and absence of relaxation in the observed time spectra.
Large standard errors on $\Lambda_{\text{BC}}^0$ and $\Lambda_{\text{BC/T}}^{0/0}$ reflect that $f(\text{Mu}_{\text{BC}}^0)$ $\simeq$ 0 at this T.
At 20~K, a fraction of muons (34~\%) are in the Mu$_{\text{BC}}^0$ state upon implantation.
They are, however, rapidly depolarized by the spin exchange interaction (77~MHz) or converted to Mu$_{\text{T}}^0$ (24~MHz).
In this sample, the conduction electron density n$_\text{e}$ at 20~K is 6.2 $\times$ 10$^{15}$ cm$^{-3}$, whereas this is 2.2 $\times$ 10$^{16}$ cm$^{-3}$ at 290~K.
This paucity and slower thermal velocity of electrons at low T results in slowing down $\Lambda_{\text{T}}^0$ and $\Lambda_{\text{T}}^{0/-}$, and providing chances for Mu$_{\text{T}}^-$ to spontaneously release one of its electrons (0.07~MHz) and go back to Mu$_{\text{T}}^0$, where the muon spin can be depolarized by the HF interaction.
This mechanism is behind the slow relaxation observed in FIG.~\ref{fig:Fig1_dark_spectra}(b).

Transition channels that involve charge carriers can be characterized by capture or scattering cross sections.
Three transition rates in Table~\ref{table:dark_results} can be described as follows:
\begin{eqnarray}
  \Lambda_{\text{T}}^0 & = & n_\text{e} v_\text{e} \sigma_{\text{T}}^0, \\
  \Lambda_{\text{T}}^{0/-} & = & n_\text{e} v_\text{e} \sigma_{\text{T}}^{0/-}, \\
  \Lambda_{\text{BC}}^0 & = & n_\text{e} v_\text{e} \sigma_{\text{BC}}^0,
\end{eqnarray}
where
$\sigma_{\text{T}}^0$ and $\sigma_{\text{BC}}^0$ are an electron scattering cross section for Mu$_{\text{T}}^0$ and Mu$_{\text{BC}}^0$ respectively, and
$\sigma_{\text{T}}^{0/-}$ is an electron capture cross section for the Mu$_{\text{T}}^0$ center.
An electron thermal velocity $v_\text{e}$ can be calculated from the equipartition theorem with the effective mass of electrons in GaAs.
The dark LF scan data were taken at seven temperatures.
For each T, we applied the fitting procedure, obtained transition rates, and calculated the $\sigma$'s
(full results can be found in Ref. \cite{SM}).
As a result, the temperature dependence of the cross sections was obtained as shown in FIG.~\ref{fig:Fig2_Mu_diagrams}(b).
Firstly, $\sigma_{\text{T}}^0$ and $\sigma_{\text{T}}^{0/-}$ are an order of magnitude smaller than $\sigma_{\text{BC}}^{0}$ in this T range.
This is a manifestation of the physical size of the Mu$_{\text{T}}^0$ and Mu$_{\text{BC}}^0$ center; 
as characterized by their HF constants,
the Mu$_{\text{T}}^0$ state is compact and huddles at the tetrahedral interstitial site,
whereas Mu$_{\text{BC}}^0$ is localized at the bond-center site but the wave function of its bound electron is extended to the crystal lattice with a higher probability on neighboring atoms.
Secondly, it is notable that $\sigma_{\text{T}}^0$ and $\sigma_{\text{T}}^{0/-}$ are at the similar values and have the same characteristic T dependence with their minimum at about 100~K.
This behavior should be contrasted with the temperature dependence of the Mu$_{\text{T}}^0$ hop rate measured in high-resistivity GaAs, which exhibited the similar behavior and took its minimum value at 90~K \cite{Kadono_GaAs_1991}.
Kadono et al. explained the behavior that, above the crossover temperature T$_\text{X}$, the system is thermally activated and phonon-assisted tunneling is predominant;
below T$_\text{X}$, incoherent quantum tunneling becomes a dominant process with decreasing temperature.
Here, we observed the same microscopic dynamics of Mu$_{\text{T}}^0$ {\it i.e.}
a higher rate of site-to-site tunneling makes the apparent scattering/capture cross sections larger.
In addition, we observed a T dependence of $\sigma_{\text{BC}}^0$, which monotonically increased with decreasing temperature.
This behavior can be attributed to the same mechanism as Mu$_{\text{T}}^0$, where the tunneling takes place from a bond center site to the next nearest site.
However, the Mu$_{\text{BC}}^0$ tunneling may require more energy than Mu$_{\text{T}}^0$ because Mu$_{\text{BC}}^0$ formation requires an increase in the Ga-As spacing \cite{Kiefl};
this could result in a higher T$_\text{X}$.

\begin{figure}
\includegraphics{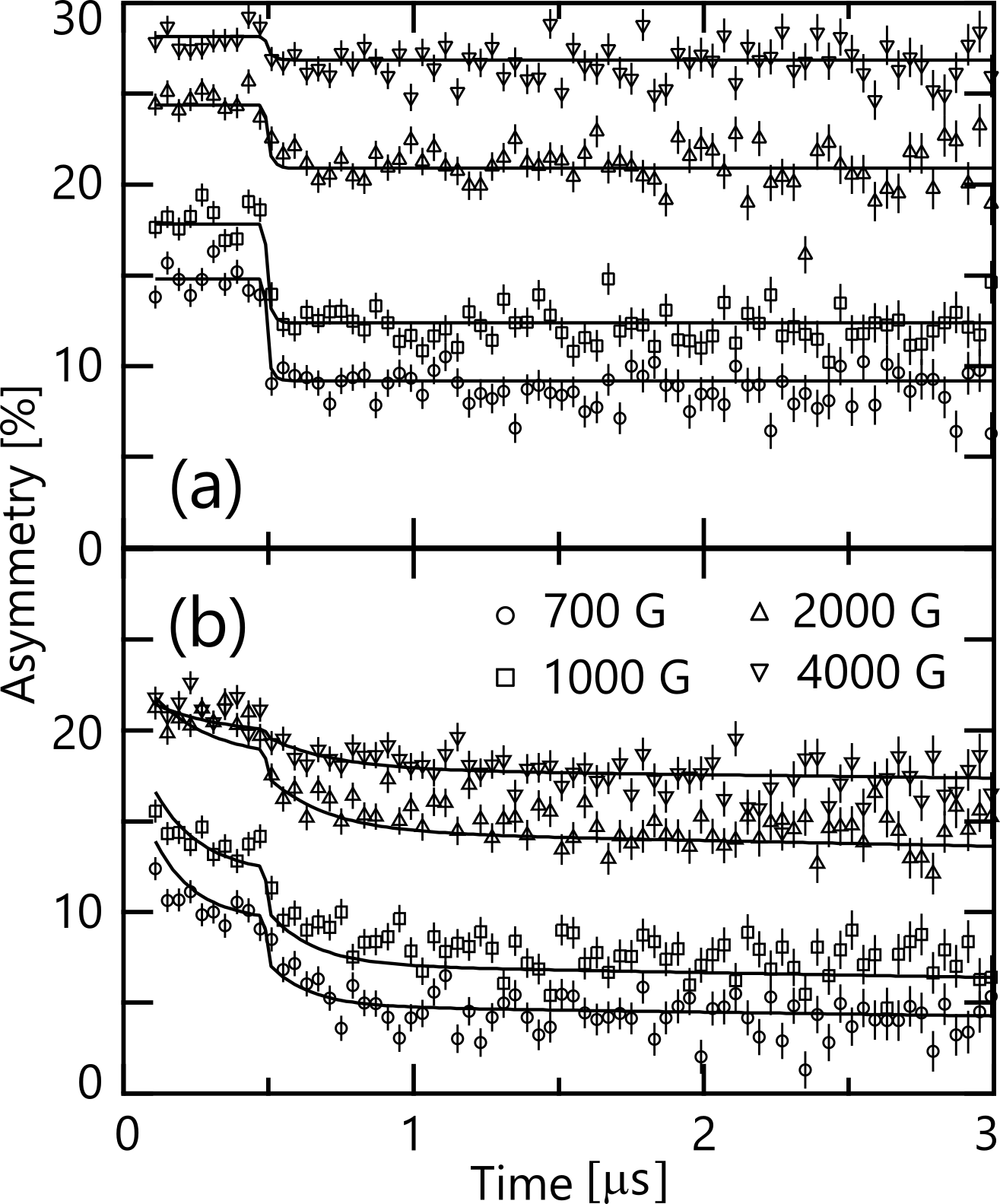}
\caption{\label{fig:Fig3_light_spectra} 
Photo-$\mu$SR time spectra measured at (a) 290~K and (b) 20~K under four representative fields.
The pump wavelength was 898 and 833~nm for (a) and (b) respectively.
The laser pulse at $\Delta$t = 0.5~$\mu$s generated $\Delta$n = 2.6 $\times$ 10$^{15}$ and 1.2 $\times$ 10$^{16}$ cm$^{-3}$ for (a) and (b) respectively.
Solid lines denote the fit results (see text).}
\end{figure}

In order to test the state exchange model and measure capture cross sections for holes, we performed photo-$\mu$SR studies on this sample.
Previous studies on GaAs reported muon spin depolarization upon carrier generation by laser light \cite{Yokoyama_PhysicsProcedia}.
Here, arrival of a laser pulse was measured at the sample position and carefully timed with respect to the muon pulse \cite{Yokoyama_RSI}.
The pulse delay $\Delta$t was set to 0.5~$\mu$s to have a dark period at the beginning of a time spectrum;
this was to confirm experimental consistency with the fully dark measurements (FIG. \ref{fig:Fig1_dark_spectra}).
Short carrier lifetimes in GaAs ($\lesssim$1~ns) mean that carriers can interact with Mu only during the laser pulse duration, which is 8~ns FWHM in the setup.
Hence, as shown in FIG.~\ref{fig:Fig3_light_spectra}, the carrier-induced depolarization is observed as a step-like change in the time spectra.
To maximize the signal, the pump wavelength needs optimization;
if the photon energy $h\nu$ is too small compared with the gap energy E$_\text{g}$, the sample becomes transparent  for the wavelength and few carriers are generated.
If, however, $h\nu$ is larger than E$_\text{g}$, all photons are absorbed near the surface and there are no free carriers generated in bulk.
Hence, as shown in Ref. \cite{SM}, a signal peak can be found in the wavelength scan, where the spatial overlap of generated carriers and implanted muons is maximized.
Since E$_\text{g}$ is dependent on temperature, this measurement for an optimum wavelength was repeated at every temperature.
By virtue of the monochromaticity of laser light, the excess carrier density $\Delta$n can be calculated from a measured photon flux
(typically, a pulse energy of $\approx$1~mJ illuminated an area of $\approx$5~cm$^2$)
and absorption coefficients \cite{Sturge}.

With these optical parameters, the photo-$\mu$SR time spectra shown in FIG.~\ref{fig:Fig3_light_spectra} were analyzed.
For the RT data [FIG.~\ref{fig:Fig3_light_spectra}(a)], the simulation included additional transition rates induced by the excess carriers {\it i.e.},
\begin{eqnarray}
  \tilde{\Lambda}_{\text{T}}^{-/0} & = & \Lambda_{\text{T}}^{-/0} + \Delta n v_\text{h} \sigma_{\text{T}}^{-/0}, \\
  \tilde{\Lambda}_{\text{T}}^{0/-} & = & \Lambda_{\text{T}}^{0/-} + \Delta n v_\text{e} \sigma_{\text{T}}^{0/-}, \\
  \tilde{\Lambda}_{\text{T}}^{0} & = & \Lambda_{\text{T}}^{0} + \Delta n v_\text{e} \sigma_{\text{T}}^0, \\
  \tilde{\Lambda}_{\text{BC}}^{0} & = & \Lambda_{\text{BC}}^{0} + \Delta n v_\text{e} \sigma_{\text{BC}}^0,
\end{eqnarray}
where the second term in Eq.~(4) describes a Mu$_{\text{T}}^-$ center capturing a hole with the cross section $\sigma_{\text{T}}^{-/0}$ and the hole thermal velocity $v_\text{h}$.
In the simulation, these additional terms were activated only during the 8-ns pulsed illumination.
Although $\Delta$n changes with time because of the pulse's temporal profile and the excess carrier decay, we assumed a constant $\Delta$n during this period for simplicity in the computation.
Similarly to FIG.~\ref{fig:Fig1_dark_spectra}, a global fit was performed on the set of LF-scan data with $\sigma_{\text{T}}^{-/0}$ as the only fit parameter;
all other parameters were adapted from the dark measurement.
Here, we assumed that free excess carriers would dominate the transitions at RT.
At 20~K, however, most of them form free excitons according to the Boltzmann distribution because the exciton binding energy in GaAs is 4.2~meV \cite{Nam}.
Hence, to analyze the data in FIG.~\ref{fig:Fig3_light_spectra}(b), $v_\text{e}$ and $v_\text{h}$ in the second term in Eq. (4) -- (7) were replaced by a free exciton thermal velocity $v_x$.
Here, the effective mass of an exciton was assumed to be a sum of the electron and hole effective mass \cite{Mattis}.
Obtained fit results are shown in FIG.~\ref{fig:Fig3_light_spectra} with with $\sigma_{\text{T}}^{-/0}$ = 3.62(4) and 3.77(7) $\times$ 10$^{-15}$ cm$^{2}$ for 290 and 20~K respectively.
The good agreement of the data points and fitting curves at both temperatures support validity of the state exchange model [FIG.~\ref{fig:Fig2_Mu_diagrams}(a)].
The enhanced cross section, $\sigma_{\text{T}}^{-/0}$ being an order of magnitude greater than $\sigma_{\text{T}}^{0}$ and $\sigma_{\text{T}}^{0/-}$, can be attributed to the Coulomb interaction between Mu$_{\text{T}}^-$ and h$^+$.
Note that $\sigma_{\text{T}}^{-/0}$ at 290 and 20~K are almost identical ---
presumably, the cross section only weakly depends on temperature because the Mu$_{\text{T}}^-$ center is known to remain static on the $\mu$SR timescale if the temperature is below 500~K \cite{Chow_Rev}.

\begin{figure}
\includegraphics{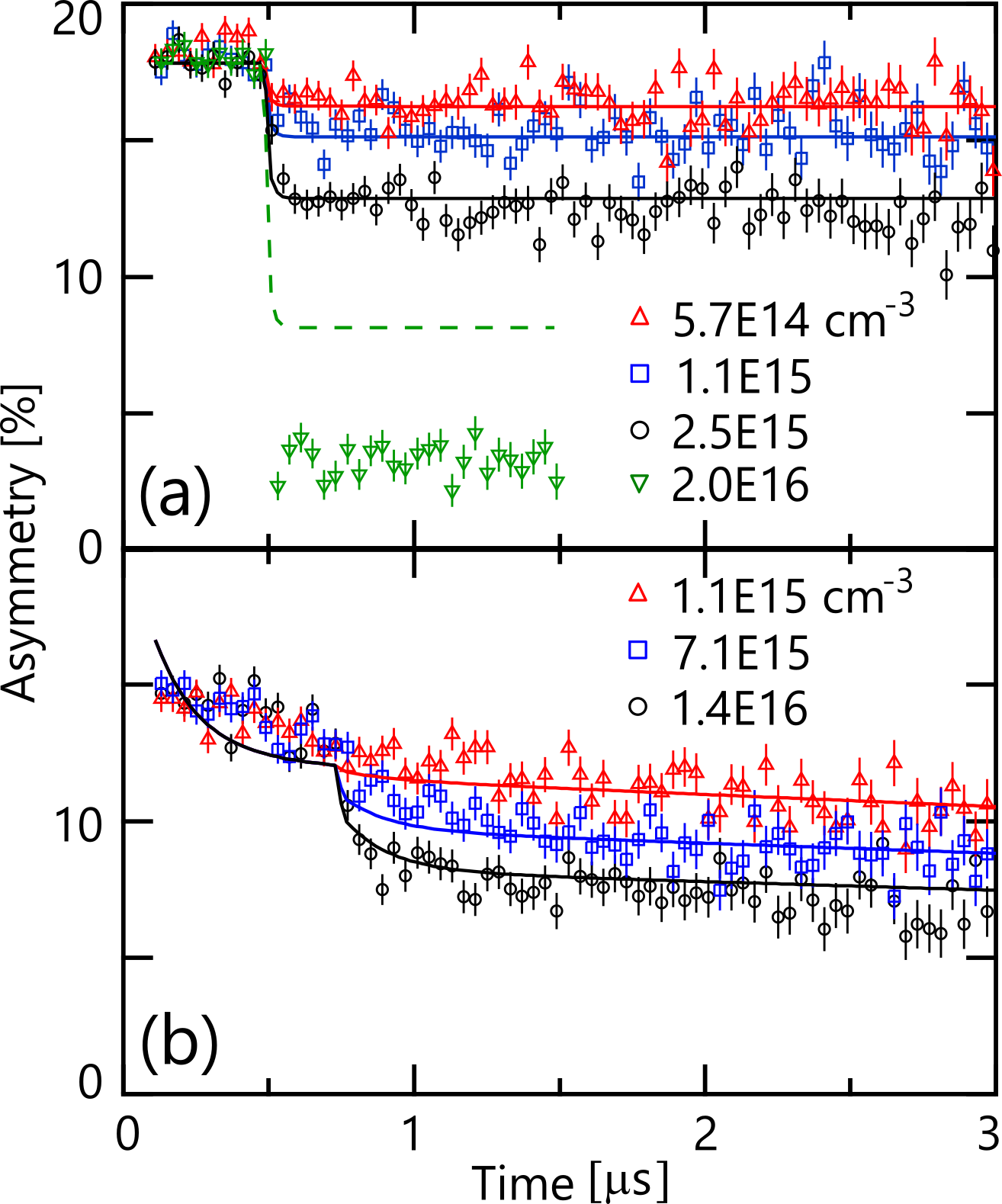}
\caption{\label{fig:Fig4_light_spectra_power}
Laser power dependent photo-$\mu$SR time spectra measured at (a) 290~K and (b) 20~K under a 1~kG LF.
Time spectra for representative $\Delta$n are shown here.
The laser pulse timing for the 20~K data was $\Delta$t = 0.75~$\mu$s.
Solid lines denote the fit results (see text).
The dashed line in Fig. (a) denotes a calculated curve with $\Delta$n = 2.0 $\times$ 10$^{16}~$cm$^{-3}$ and the obtained parameters.
}
\end{figure}

The state exchange model was further tested by measuring the power dependence {\it i.e.} the excess carrier density.
Figure~\ref{fig:Fig4_light_spectra_power} shows the $\mu$SR time spectra for representative excess carrier densities.
Here, a LF of 1000~G was applied to the sample to decouple stray fields in the instrument and local fields by crystalline defects and impurities (but not strong enough to completely decouple the Mu HF interaction).
A set of calibrated neutral density filters were used to change the pump laser power.
Similarly to the fitting procedure in FIG.~\ref{fig:Fig3_light_spectra}, global fitting was applied to a set of the $\Delta$n-dependent data with $\sigma_{\text{T}}^{-/0}$ as the only fit parameter.
The best fit gave $\sigma_{\text{T}}^{-/0}$ = 3.22(4) and 1.81(3) $\times$ 10$^{-15}~$cm$^{2}$ for 290 and 20~K respectively.
This reasonably good agreement with the LF results, once again, supports correctness of the model at this level of excess carrier density.
Indeed, fit quality for the RT data gets worse if it includes data with $\Delta$n $>$ 2.5 $\times$ 10$^{15}~$cm$^{-3}$, where the model seems to require additional processes.
As shown in Eq.~(4) -- (7), our model included contribution by free holes only in $\tilde{\Lambda}_{\text{T}}^{-/0}$.
However, if $\Delta$n is sufficiently high, other Mu reactions with the holes can take place {\it e.g.}
Mu$_{\text{T}}^0$ + h$^+$ $\rightarrow$ Mu$_{\text{T}}^+$ + e$^-$ $\rightarrow$ Mu$_{\text{T}}^0$;
this cyclic process is equivalent to adding an extra term to $\tilde{\Lambda}_{\text{T}}^{0}$.
Indeed, as shown in FIG.~\ref{fig:Fig4_light_spectra_power}(a), the model fails to simulate the high $\Delta$n data (dashed line).
With lower pumping, reactions by excess electrons are predominant in the exchange processes because of their faster thermal velocity;
the contributions by slower holes are negligible in most transition channels except the capture by Mu$_{\text{T}}^-$.
In contrast, as shown in FIG.~\ref{fig:Fig4_light_spectra_power}(b), the deviation from the model was not observed at 20~K, even though a hole should be located at the vicinity of an electron because of the exciton formation.
This could be attributed to the wave function of the bound electron spatially more spread out than the hole's because of its lighter effective mass.
The larger wave packet results in a higher probability for electrons interacting with the Mu centers, which, in turn, dominates the transition mechanism.

In summary, we identified the muonium state exchange mechanism in n-GaAs by fitting the $\mu$SR data with the density matrix simulation.
The photo-$\mu$SR measurements played a key role in testing the model and studying the system response to carrier generation.
The model turned out to be rather simple, especially at RT, where the Mu$_\text{T}$ center is the predominant species.
Our studies not just agreed with the previous studies, but also provided new insight into underlying mechanisms and capture/scattering cross sections of the Mu states.
The Mu behavior learned from this study can be applied to some of the cases with hydrogen, if the difference in their mass is correctly considered.
For instance, hydrogen atoms in a precursor state probably form the paramagnetic states and go into the state exchange mechanism before converted to their final stable states.
Their scattering cross sections (and transition rates), however, can be different from our measurements because of the difference in zero-point energy.
Looking to future development, firstly, this study holds the key to understanding the muon signal upon generating electronic spin polarization in n-GaAs \cite{Yokoyama_PhysicsProcedia}.
Yokoyama et al. performed a photo-$\mu$SR experiment with optical spin orientation in n-GaAs and found that the amount of optically induced relaxation depended on the helicity of circularly polarized light {\it i.e.} the direction of the conduction electron spin polarization.
With electronic spins polarized in the system, we can expect that the electron capture rate of Mu$_{\text{T}}^0$ to form Mu$_{\text{T}}^-$ should obey the Pauli's exclusion principle and depend on availability of electrons with a right spin.
The state exchange model elucidated in this study lays a foundation for understanding this phenomenon, which can lead to $\mu$SR applications to spintronics studies.
Secondly, we demonstrated that the simulation method and illumination technique were a powerful combination to study a underlying mechanism of $\mu$SR data.
Although Mu dynamics in silicon (and germanium) were extensively studied \cite{Chow_Rev}, the picture may not be fully conclusive yet.
An application of the method can provide new information with a full state exchange model.
This can be the case for other semiconductor systems, such as SiC, where Mu is formed and undergoes state exchange in the timescale of a muon lifetime.

This work was carried out using a beamtime allocated by the STFC ISIS Facility \cite{DOI}.
Support is acknowledged from the Texas Research Incentive Program (P.W.M., R.L.L.) and the NMU Freshman Fellows Program (M.R.G.).
K.Y. and J.S.L. would like to thank Alan J. Drew of Queen Mary University of London for his effort and support on developing the photo-$\mu$SR setup in the HiFi spectrometer.
Finally, we are grateful for the assistance of a number of technical and support staff in the ISIS facility.


\end{document}